\documentclass[12pt]{article}
\pdfoutput=1
\usepackage{a4wide,amssymb,cite}
\usepackage{epsfig,graphicx}
\usepackage[usenames,dvipsnames]{color}
\usepackage{slashed}

\newcommand{\beq}{\begin{equation}}
\newcommand{\eeq}{\end{equation}}
\newcommand{\bea}{\begin{eqnarray}}
\newcommand{\eea}{\end{eqnarray}}

\begin{document}

\color{black}

\begin{flushright}
KIAS-P13045, FTUV-13-47, IFIC-13-73
\end{flushright}

\vspace{1cm}
\begin{center}
{\Large\bf\color{black} Coleman-Weinberg Inflation in light of Planck}\\
\bigskip\color{black}\vspace{1.5cm}{
{\bf Gabriela Barenboim$^1$, Eung Jin Chun$^2$ and Hyun Min Lee$^{3}$} \vspace{0.5cm} } \\[7mm]

{\it $^1$Departament de F\'{\i}sica Te\`orica and IFIC, Universitat de Val\`encia-CSIC, \\ E-46100, Burjassot, Spain.
} \\ {\it $^2$School of Physics, Korea Institute for Advanced Study, Seoul 130-722, Korea.  }\\
{\it $^3$Department of Physics, Chung-Ang University, Seoul 156-756, Korea. }
\end{center}
\bigskip
\centerline{\large\bf Abstract}
\begin{quote}
We revisit a single field inflationary model based on Coleman-Weinberg potentials. We show that in
small field Coleman-Weinberg inflation, the observed amplitude of perturbations needs
an extremely small quartic coupling of the inflaton, which might be a signature  of radiative origin.
However, the spectral index obtained in a standard cosmological scenario turns out to be outside
the 2$\sigma$ region of the Planck data. When  a non-standard cosmological framework is invoked,
such as brane-world cosmology in the Randall-Sundrum model, the spectral index can
be made consistent with  Planck data within $1\sigma$, courtesy of the modification in the evolution of the Hubble
parameter in such a scheme. We also show that the required inflaton quartic coupling as well as a phenomenologically
viable $B-L$ symmetry breaking together with a natural electroweak symmetry breaking can arise dynamically in a generalized $B-L$ extension of the Standard Model where the full potential is assumed to vanish at a high scale.

\end{quote}

\thispagestyle{empty}

\normalsize

\newpage

\setcounter{page}{1}

\section{Introduction}

The Hot Big-Bang Model and General Relativity successfully explain the thermal history of our Universe since its very
first nanoseconds of existence up to now. However they left unexplained crucial issues such as the horizon, flatness,
and monopole problems. Inflation is the most elegant and simple explanation to address them and therefore no modern
cosmological model lacks a prudential stage of inflation. Inflation basically assumes the existence of a period of
exponential growth of the scale factor which essentially wipes out any trace of curvature, dilutes unwanted relics and
leaves the Universe in a highly symmetric state.

The simplest scenario where this picture can be realized is based on a single scalar field (called inflaton) with a
nearly flat potential. The quantum fluctuations of the inflaton would be then responsible for the  tiny  temperature
anisotropies observed in the Cosmic Microwave Background (CMB).
Basically an  inflationary theory must fulfill two requirements to agree with experimental observations: (i) it has to
provide sufficient inflation, i.e., the inflationary potential  must drive an increase on the scale factor of 50 -- 60
e-folds (the precise number depends on details of the particular inflaton model) in order to describe the thermal
equilibrium  observed in the CMB at least for the scales of interest; (ii) the size of the quantum fluctuations of the
inflaton leading to the power spectrum of primordial curvature perturbations, $A_s$, should be at the appropriate
level while the spectral index $n_s$ must agree with observations. Because of these ``mild" requirements most simple
inflation models were given a death blow once the Planck mission released its high-precision data \cite{planck13}. The
Planck result, combined with the WMAP large-angle polarization  measurements, requires the  two observables $n_s$ and
$A_s$ of curvature perturbations to be \beq n_s = 0.9603 \pm  0.0073\,, \quad A_s = 2.196^{+0.051}_{-0.060} \times
10^{-9} \eeq at  a scale $ k_* = 0.05$ Mpc$^{-1}$ which rules out exact scale-invariance at more than  5$\sigma$.
Analogously, the tensor-to-scalar ratio $r$  is bounded to be $ r < 0.11 $ at 95 \% CL, for the same scale $k_*$.
These constraints are already powerful enough to rule out or strongly disfavour the most popular and simple
inflationary potentials.

Consequently, simple and well motivated inflationary potentials are distinctively welcomed. Among these, a special
place should be given to Coleman-Weinberg (CW) type of potentials \cite{CW} as they not only arise naturally but are
unavoidable when loop corrections are taken into account. Inflation with a CW potential has been suggested at an early
stage of the inflation theory formulation, and studied extensively, in particular, in association with  grand
unification theories \cite{cwi-old}.

\medskip

In this paper, we revisit the CW inflation accounting for cosmological observations in a scheme where the inflaton
potential has a dynamical origin, naturally arising from quantum corrections. As we will see later, the quartic
coupling of the CW potential is proportional to the amplitude of primordial perturbations and thus has to be extremely
small ($\sim 10^{-14}$) which might point towards a radiative origin.
An attempt to generate a phenomenologically viable scalar potential starting from a vanishing initial condition of the full scalar potential at a cutoff scale has been worked out in the context of the $SU(3)_c \times SU(2)_L \times U(1)_Y \times U(1)_X$ model where $U(1)_X$ is a generalized $B-L$ gauge symmetry \cite{chun13}.
It has been shown that a Standard Model (SM) Higgs potential consistent with recent LHC data \cite{lhc} can be generated radiatively once the $B-L$ symmetry is broken appropriately by the usual CW mechansim.
One of the distinctive features of this scheme is that the quartic coupling of the $B-L$ scalar
arises due to its coupling to right-handed neutrinos in a similar way as
the SM scalar quartic coupling arises due to the top Yukawa coupling.

We find that in a general CW inflation model, the number of e-folding required for solving the horizon problem cannot
be made consistent with the observed spectral index. That is, the spectral index resulting from getting enough number
of e-folds  to take care of the horizon problem, can only fit within the $3\sigma$ range of the Planck data for a
symmetry breaking scale larger than $10^{15}$ GeV. In particular, when using a value of the symmetry breaking scale
favoured for providing natural electroweak symmetry breaking in a {\rm generalized} $B-L$ extension of the SM, the resulting spectral
index is well outside the $5\sigma$ range of the Planck data. This motivates us to consider the CW inflation in a non-standard cosmological scenario,  a typical example of which is  brane  cosmology in a higher dimensional space
such as the Randall-Sundrum model \cite{rsmodel,maartens99}. In this case, the extra dimension does not evolve in time while
the evolution of four-dimensional spacetime  is described by a modified Friedman equation, where $H^2\propto \rho^2$
in the large energy density limit. Thus, the effective inflaton energy increases by a factor, $V_0/\Lambda$, with
$V_0$ being the original inflaton energy and $\Lambda$ being the brane tension, which makes the CW inflation fully
compatible with the observations for an appropriate choice of the brane tension. A remarkable feature of the CW inflation
on the brane is that the correlations between the observables are kept the same as in the standard CW inflation.

The paper is organized as follows. We begin by revisiting  CW inflation in standard cosmology in Section 2. In
Section 3 the setup is extended to  brane cosmology and the inflationary observables are calculated for the
modified Friedman equation. We show in Section 4 how the required quartic coupling and symmetry breaking scale
for the CW inflaton potential can be generated in the context of the $B-L$ extension of the SM
which is motivated by  the explanation of the neutrino masses and mixing.
Finally, conclusions are drawn in Section 5.

\section{CW inflation in standard cosmology}

A general CW potential evaluated at the renormalization scale $Q = \langle \phi \rangle = v_\phi$ takes a rather
simple form: \bea \label{Apot} V(\phi) = A \phi^4 \left( \log\left(\frac{\phi}{v_\phi}\right) - \frac{1}{4} \right) +
\frac{A}{4} v_\phi^4 \eea which satisfies $V'(v_\phi)=0$ and $V(v_\phi)=0$. Here $A=A(v_\phi)$ is determined by the
beta function for the scalar quartic coupling defined at $Q=v_\phi$, which is a function of gauge, Yukawa and other
scalar couplings of the inflaton field $\phi$. A detailed form of the beta function and the CW potential applied to
the $B-L$ gauge symmetry will be discussed later. Depending on the values of $A$ and $v_\phi$, the inflaton  can have
small or large values compared to the Planck scale during observable inflation. In the latter case inflation mimics
chaotic inflation.
We will be interested in the scenario where the inflaton starts its journey from small values of
the field, $\phi/v_\phi<1$, in the flat region of the CW potential as drawn in Fig.~1.
There is a recent overview on various inflation models including the CW type inflation \cite{martin}.

\begin{figure}[t]
\centering
\includegraphics[width=8cm]{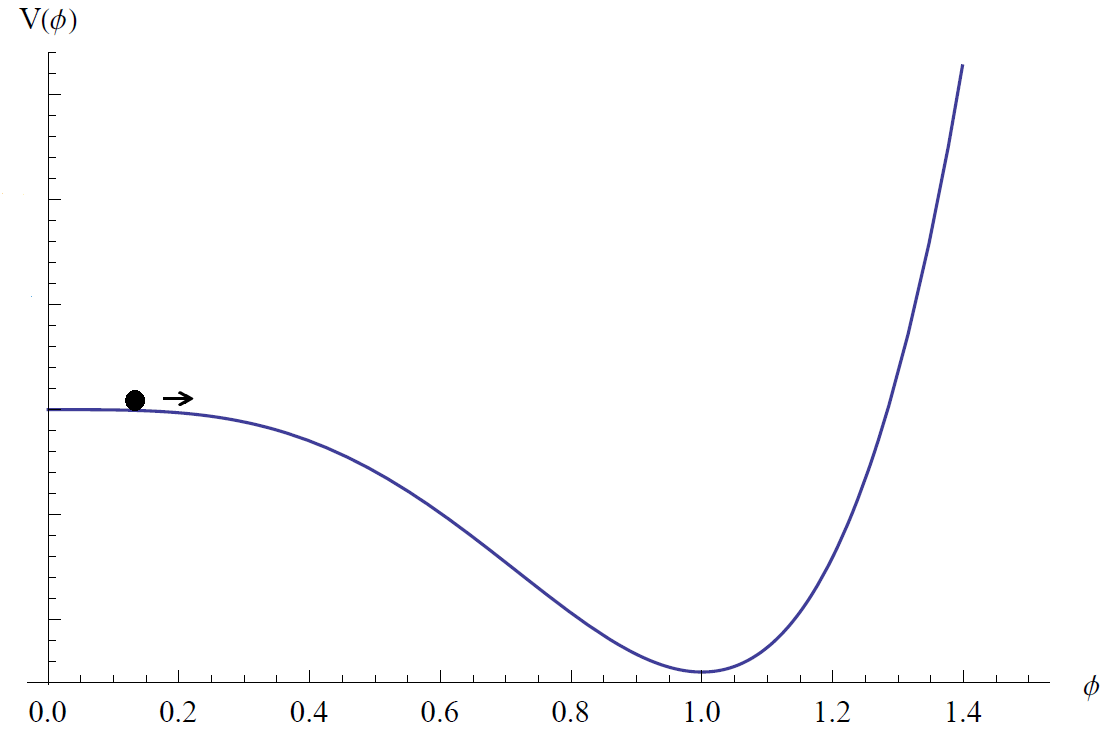}
\caption{The CW potential realizing a small field inflation. } \label{fig:cw-pot}
\end{figure}

During inflation,  the inflaton rolls down towards the minimum of its potential, evolving according to \beq
\ddot{\phi} + 3 H \dot{\phi} + \partial V/\partial\phi =0 \eeq where the Hubble rate $H$ is given by \beq H^2\; =
\;\frac{1}{3 M_{P}^2} \; \left[ \frac{\dot{\phi}^2}{2} + V(\phi) \right]. \eeq Here $M_P \approx 2.44 \times 10^{18}$
GeV is the reduced Planck mass. If the field is moving along a region where the potential is so flat that the
evolution of the
 field is friction dominated, (what goes in the literature under the name of  slow-roll approximation) the equation of
 motion
is basically given by \beq 3 H \dot{\phi} + \partial V/\partial\phi \approx 0. \eeq Within this approximation, the
number of e-folds of inflation generated since the modes that are entering the observable Universe now left the
horizon (at $\phi_* $) till the end of inflation (at $\phi_f$) is given by \bea N(\phi_*) =  \;-\frac{1}{M_{P}^2} \;
\int_{\phi_*}^{\phi_f} \frac{V(\phi)}{V^\prime(\phi)} d\phi \eea where $V^\prime(\phi) = \partial V/\partial\phi $ and
$\phi_f$ is the value of the field at which
 inflation stops. This happens when
\beq \epsilon(\phi_f) \equiv \frac{M_{P}^2}{2} \left[ \frac{V^\prime(\phi_f)}{V(\phi_f)} \right]^2 = 1 \,. \eeq At
this time, quantum fluctuations on the scale observed today were produced too and its size is given by \bea {\cal
P_R}(k_*) = {1\over 24 \pi^2} {V(\phi_*) \over M_P^4 \, \epsilon(\phi_*) } \eea while the spectral index of these
density perturbations and its dependence on the scale  read \bea n_s - 1  & = & - 6 \epsilon(\phi_*) + 2 \eta(\phi_*)
\,, \\ \nonumber \\ \frac{d n_s}{d \ln k} & =&  16 \epsilon(\phi_*) \eta(\phi_*) - 24 \epsilon(\phi_*)^2 - 2
\xi^2(\phi_*) \,. \eea Here, $\eta$ and $\xi^2 $ are the second  and third slow-roll parameters that can be expressed
in terms of the inflaton potential as \bea \eta \;=\;  M_P^2 \frac{V^{\prime \prime}}{V}   \;\; \;\; \mbox{and} \;\;
\;\; \xi^2 \;=\; M_P^4 \frac{ V^{\prime \prime \prime} \; V^\prime}{V^2}. \eea Besides the observed  density (scalar)
perturbations, the inflaton generates the yet unobserved gravitational waves or tensor perturbations. Normally, the
tensor amplitude is expressed in terms of the tensor/scalar ratio as \beq r \equiv \frac{\cal P_T}{\cal P_R} = 16
\epsilon(\phi_*) \,. \eeq

Expressed in terms of the parameters of the CW potential (\ref{Apot}), the above inflation quantities are given
explicitly by
\bea
\epsilon &=& 8  \left( \frac{ 4M_{P}}{ v_{\phi} }\right)^2 \; \left(\frac{\phi}{v_\phi }\right)^6
\; \ln^2\left(\frac{\phi}{v_\phi}\right) \label{eq:eps} \\
\eta &=&  \left( \frac{ 4 M_{P}}{ v_{\phi} } \right)^2 \;
\left(\frac{\phi}{v_\phi }\right)^2 \;\left( 3 \ln\left(\frac{\phi}{v_\phi}\right) +1 \right) \label{eq:eta} \\
\xi^2 &=&  \left( \frac{ 4 M_{P}}{ v_{\phi}} \right)^4 \; \left(\frac{\phi}{v_\phi }\right)^4 \;
\ln\left(\frac{\phi}{v_\phi}\right) \; \left( 6 \ln\left(\frac{\phi}{v_\phi}\right) + 5 \right)  \label{eq:xi} \\
N &=&  \left( \frac{v_\phi}{ 4 M_{P}} \right)^2 \; \left( \mbox{Ei}\left[ -2 \ln\left(\frac{\phi}{v_\phi}\right)
\right]-\mbox{Ei}\left[ -2 \ln\left(\frac{\phi_f}{v_\phi}\right) \right] \right) \label{eq:N} \\
{\cal P_R}(k) &=& {A \over 3 \pi^2} \left( v_\phi \over 4 M_P \right)^6 \;
\left(\frac{v_\phi }{\phi}\right)^6 \; \frac{1}{\ln^2\left(\frac{\phi}{v_\phi}\right)} \label{eq:PR}
\eea
where Ei stands for the exponential integral.

\medskip

From the equations above, one can see that for slow-roll inflation to take place, a field value $\phi_*$ at horizon
exit can be always found for  {\it any} choice of $v_\phi$
and it lays in the region of $\phi_*/v_\phi \ll 1$ as the slow-roll parameters have
a large factor of $(4 M_P/v_\phi)^2$. This leads to
$|\epsilon| \ll|\eta| \approx \sqrt{3|\xi^2|/2}$.
As a consequence, small field CW inflation predicts remarkable correlations:
\bea
 N &\approx& {3\over 1-n_s} \,,  \label{Nns}\\
 \frac{d n_s}{d \ln k} &\approx& -\frac{1}{3}(1-n_s)^2 \,, \label{ns-run}
\eea
independently of the precise values of $v_\phi$ and $\phi/v_\phi$.
Similarly, the power spectrum of perturbation is
\bea {\cal P_R}(k)& \approx& A  {72 \over \pi^2}  \,{ \left|
\ln(\phi/v_\phi) \right| \over (1-n_s)^3 }, \label{PR}
\eea
which is insensitive to $\phi/v_\phi$. Considering the central value
of  $n_s= 0.96$  and ${\cal P_R} \approx 2.2\times10^{-9}$ measured by the Planck mission at the pivot scale $k_* = 0.05$ Mpc$^{-1}$ \cite{planck13}, one gets $N\approx  75$, and $A\sim 10^{-14}$ mildly depending on
$\phi_*$. The tensor-to-scalar ratio given by \bea r&\approx& \frac{16}{27} \left(\frac{v_\phi}{4M_P}\right)^4
\,\frac{(1-n_s)^3}{\left|\ln(\phi/v_\phi)\right|} \eea is severely suppressed by a factor of $(v_\phi/M_P)^4$ while
being insensitive to $\phi$.

\medskip

Such large $N$ and tiny $A$ do not fit well with the number of e-folding required to solve the horizon problem for
scales that have just entered/are entering the current horizon. The condition for such a thing to happen is \beq
1=\frac{(aH)_*}{(aH)_0} = \frac{a_*}{a_{\rm end}}\cdot \frac{a_{\rm end}}{a_{\rm reh}}\cdot \frac{a_{\rm
reh}}{a_0}\cdot \frac{H_*}{H_0}. \eeq which translates into \bea \label{Nstar}
 N_* &=& \frac{1}{3}\ln\left(\frac{\rho_{\rm reh}}{\rho_{\rm end}}\right)+\frac{1}{4}
 \ln\left(\frac{\rho_{0r}}{\rho_{\rm reh}}\right)+\frac{1}{2}\ln\left(\frac{\rho_*}{\rho_0}\right) \nonumber \\
 &\simeq& 61-\ln\left(\frac{10^{16}{\rm GeV}}{V^{1/4}_*}\right)+\ln\left(\frac{V^{1/4}_*}{V^{1/4}_{\rm end}}\right)
 -\frac{1}{3}\ln\left(\frac{V^{1/4}_{\rm end}}{\rho^{1/4}_{\rm reh}}\right)
 \eea
where $\rho_{0(r)}, \rho_{\rm reh}, \rho_{\rm end}$ are (radiation) energy densities at present, the end of
reheating and the end of inflation, respectively, and $\rho_*$, $V_*$ are energy density and inflaton potential
evaluated at horizon exit for the pivot scale $k_* = 0.05$ Mpc$^{-1}$.
Note that one can take $V_* \simeq  V(0) \simeq V_{\rm end}$ as a good approximation in
most cases. For the CW picture of inflation to work, the number of e-folds $N \big|_{\phi_*}$ (\ref{eq:N}) derived from the CW
potential should be able to reproduce the number $N_*$ (\ref{Nstar}).
This condition together with the COBE
normalization, ${\cal P_R} \simeq 2.2\times 10^{-9}$, and the CW inflation property, $n_s \simeq 1 + 2 \eta$, uniquely fixes the values of the three parameters, $A$, $v_\phi$ and $\phi/v_\phi$ of the CW potential.

\begin{figure}[t]
\centering
\includegraphics[width=7.5cm]{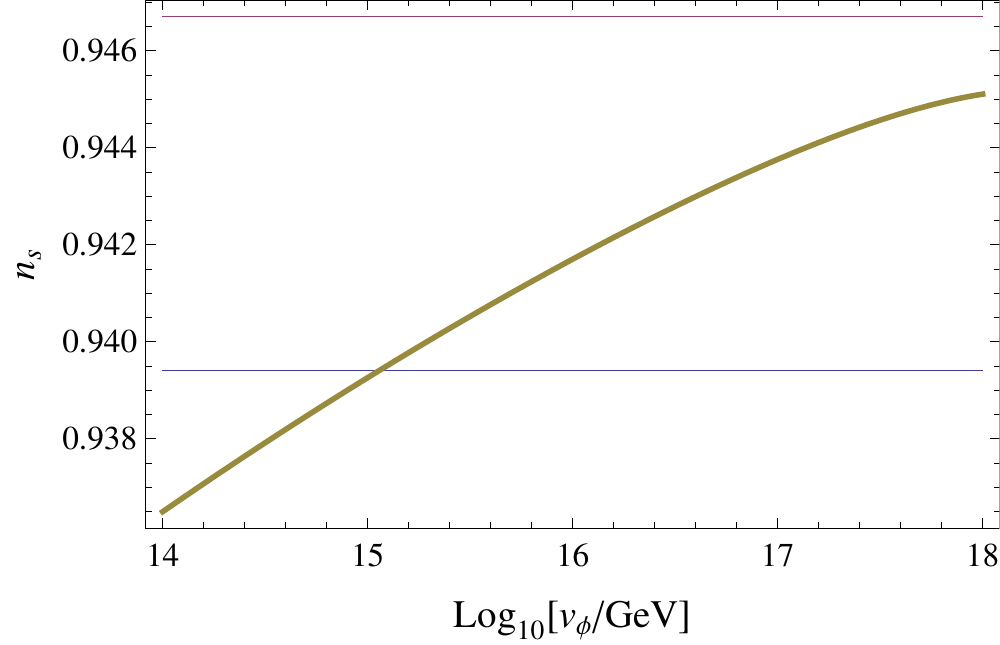}\hspace{1ex}
\includegraphics[width=8.0cm]{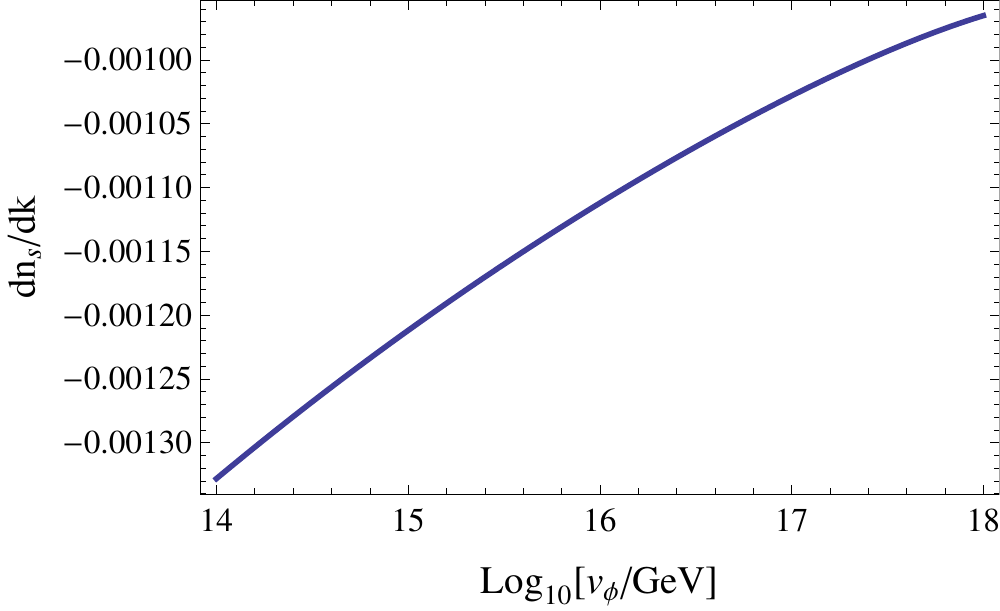}
\caption{The spectral index $n_s$ and its running as a function of $v_\phi$ in the standard CW inflation.
Two horizontal lines in the left panel show $2\sigma$ and $3\sigma$ lower bounds measured by Planck, respectively. } \label{fig:st1}
\end{figure}

The obtained spectral
index $n_s$ for a given symmetry breaking scale $v_\phi$ is shown in Fig.~\ref{fig:st1}. The largest possible value that
$n_s$ can take is about 0.945 for $v_\phi$ close to $M_P$ which is below the $2\sigma$ range favoured
by the Planck data. Only a $3\sigma$ compatibility can be achieved for $v_\phi \gtrsim 10^{15}$ GeV,
disfavoring CW inflation at the $2\sigma$ level. The running of the spectral index is presented in the right panel of Fig.~\ref{fig:st1} which shows a strong correlation between the spectral index and its running as discussed before.
The obtained running has to be compared with  Planck's result
\beq
 {d n_s \over d \ln k} = -0.013 \pm 0.009
\eeq
at 68\% CL. It is clear that the results have no definite impact on this class of models, but are
are bounded to test them in the near future as the precision on both measurements, the spectral index and its running
will be significantly improved.

\begin{figure}[t]
\centering
\includegraphics[width=8.0cm]{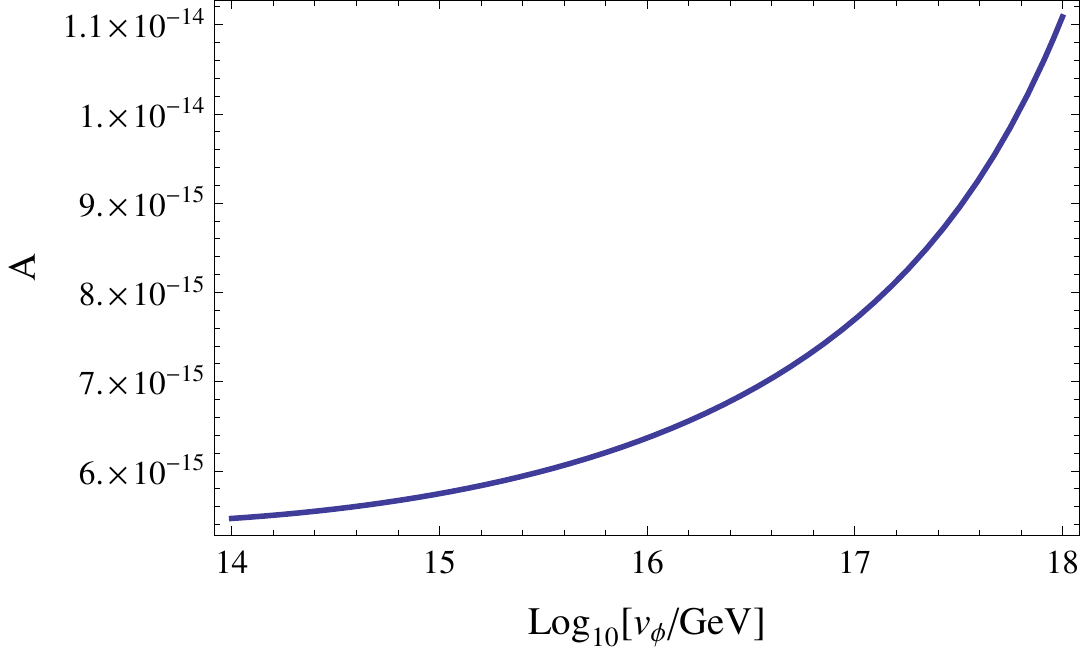}\hspace{1ex}
\includegraphics[width=7.7cm]{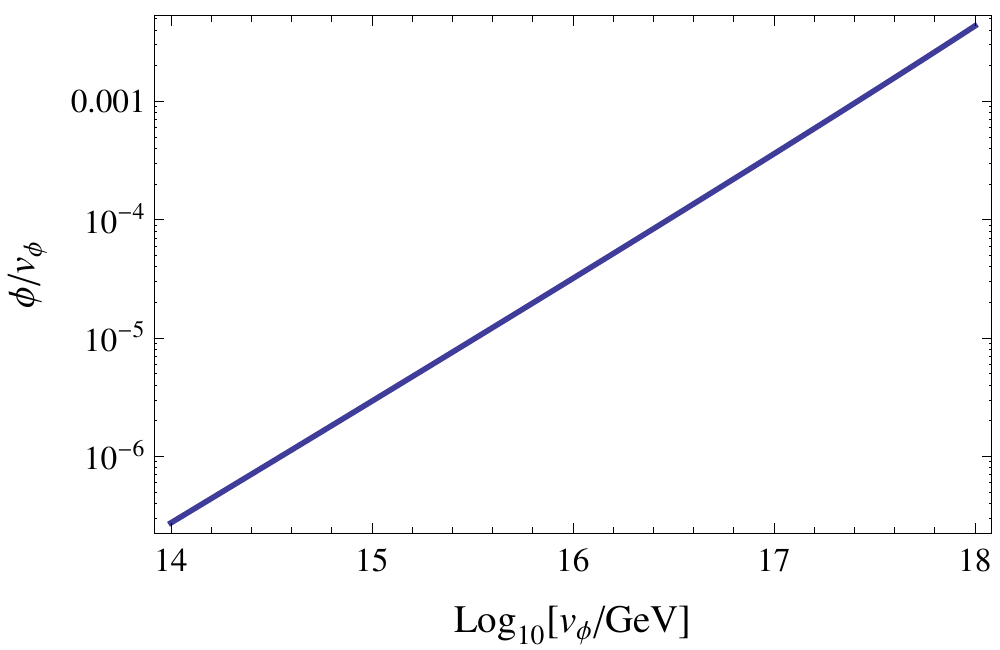}
\caption{ The correlations of $A$ and $\phi/v_\phi$ with $v_\phi$ for the CW inflation in the standard cosmology }
\label{fig:st2}
\end{figure}

In Fig.~\ref{fig:st2}, we present the values of $A$ and $\phi/v_\phi$ for given $v_\phi$ showing
a correlation between $A$ and $\phi/v_\phi$.  One finds that smaller $v_\phi$ require smaller $A$ and $\phi/v_\phi$,
but one gets $A\sim 10^{-14}$ which is  rather insensitive to $v_\phi$. As mentioned earlier, we find
$\phi/v_\phi \ll 1$ for all the region of $v_\phi < M_P$.

\medskip

The strong tension of the CW inflationary setup with  Planck data arises due to the fact  that
$V_*$ cannot be made large enough to enhance $N_*$ as a tiny quartic coupling $A$ is also required
by the size of perturbations, and even though one can have $v_\phi$ as large as $M_P$
it is not enough to overcome the strong $A$ suppression. The tension is aggravated even further
if one wants to have lower symmetry breaking scales in the CW mechanism.
This problem may be avoided in a non-standard cosmological scenario where the  relation
(\ref{Nstar}) is altered in a radical way.  Most of the modifications we can get within
the standard framework, like an earlier period of matter domination, work in the opposite direction,
i.e. they lower the number of e-folding needed to solve the horizon problem.

\section{CW inflation on the brane}

For a drastic change in the thermal history of the universe we can immerse ourselves into a
brane world scenario where
we live in a brane embedded in a higher dimensional Universe. Within this scheme,  the stress-energy momentum in the
bulk can take different forms, depending on the specifics of the model. By an appropriate choice of the boundary
conditions,  the non standard behaviour of the Universe on the brane we are looking for, can be easily achieved. Even
more, the Hubble rate itself on the brane changes rather drastically taking the form
\beq \label{Hnonst}
 H^2 = {1\over 3 M_P^2} \rho \left( 1 + {\rho^n \over M_B^{4n} } \right) + \frac{C}{a^4}
\eeq
where $M_B$ denotes a certain scale below which the cosmological evolution follows the standard form. A useful
model to illustrate these effects is the brane world cosmology of the Randall-Sundrum(RS) model \cite{rsmodel}    in which the main
correction is the term quadratic in the density, i.e.,
\bea H^2={1 \over 3M_P^2}\rho\left( 1+ {\rho\over 2
\Lambda}\right),
\eea
and the new scale $M_B$ is given by the brane tension $\Lambda=M^4_B/2$ satisfying $\Lambda =
\sqrt{-6 \Lambda_{\rm bulk} M_5^3}$ where $\Lambda_{\rm bulk}$ is the bulk cosmological constant and $M_5$ is the 5-dimensional Planck
scale  \cite{maartens99}.

During the inflationary stage the energy momentum tensor on the brane is dominated by the scalar field, which is
confined to it and therefore still evolves as Eq.\ (3), as on the brane $\nabla^\nu T_{\mu\nu}=0$ holds.
The condition to sustain a period of inflation is now
\bea p < -2/3 \rho
\eea
for $\rho \gg \Lambda$ and  when the energy density is
dominated by the scalar potential
\bea H^2 &\simeq&  \left({1 \over 3M_P^2}\right) V\left[
1+{V\over2\Lambda} \right] .
\eea
In this regime  the slow-roll parameters become
\bea \epsilon_B &=& \epsilon \cdot { 1 + 2 \tilde V \over (1 + \tilde V)^2}\,, \\
\eta_B &=& \eta \cdot{ 1\over 1+ \tilde V}\,, \\
\xi^2_B &=& \xi^2 \cdot{ 1 \over (1 + \tilde V)^2}\,,
\eea
where we have defined $\tilde V \equiv V/M_B^4$.

Note that all the slow-roll parameters recover their standard forms for $\tilde V=0$, but are suppressed by
$1/\tilde V$ in the limit of $\tilde V \gg1$.
While the spectral index take the same form: $n_s = 1 -6 \epsilon_B + 2
\eta_B$, its running is given by
\beq
 {d n_s \over d \ln k} = -2 \xi_B^2 + 16 \epsilon_B \eta_B -24 \epsilon_B^2 \cdot
{1 + 3 \tilde V + 3 \tilde V^2 \over (1 + 2 \tilde V)^2}
\eeq
The power spectrum of primordial quantum fluctuations
and the tensor-to-scalar ratios are also modified to turn into
\bea {\cal P}_{{\cal R},B}(k) &=& {\cal P_R}(k) \cdot
(1+\tilde V)^3 \\ r_B &=& 16 \epsilon_B {1\over 1 + 2 \tilde V} \,,  \label{rb}
\eea
The number of e-folding during
inflation is also changed to
\beq N_B = - {1\over M_P^2} \int^{\phi_f}_{\phi_*} {V \over V'}
\left( 1 + \tilde V\right) \, d\phi \;.
\eeq

\begin{figure}[t]
\centering
\includegraphics[width=10cm]{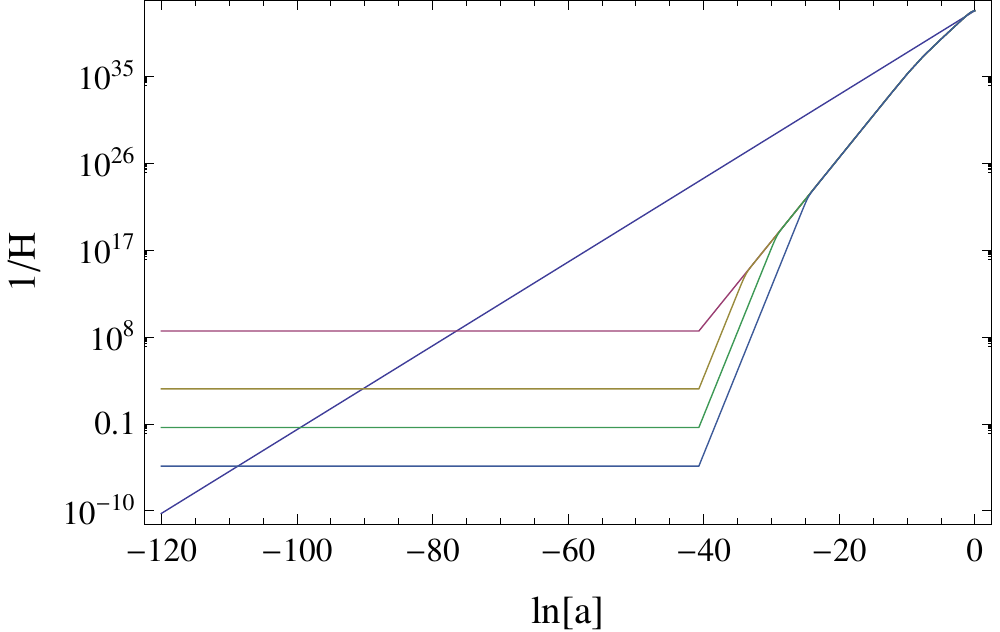}
\caption{Evolution of the horizon size in a brane cosmology with $n=1$ and
 $V_*^{1/4} = 10^5$ GeV in terms of $\ln(a)$ with the normalization of $a_{\rm now}=1$.
 The four lines correspond to $M_B =$ $\infty$ (standard),  $10^2$, $1$
 and $10^{-2}$ GeV from the top, respectively, showing how the e-folding number
 after the horizon exit of the current horizon scale changes.
} \label{fig:N-RS}
\end{figure}

The condition for solving the horizon problem now gets  an extra term which leads to an
additional number of e-folding for the scales of interest ($k_* = 0.05$ Mpc$^{-1}$) depending on $\tilde V$:
\beq
\label{NB} N_{B,*} \approx 61 + {1\over2} \ln\left(1+ \tilde V\right) - \ln\left( 10^{16} \, \mbox{GeV} \over
V_*^{1/4} \right) - {1\over3} \left(  V_*^{1/4} \over \rho_{reh}^{1/4}  \right) \,.
\eeq
This increase can be easily understood by seeing in Fig.~\ref{fig:N-RS} how steeper the Hubble radius change is,
once the new scenario kicks in.
The lower the scale associated with the new scenario (the brane tension) the larger the required number of e-folds
for a fixed value of the potential at the end of inflation, i.e., for a fixed reheating temperature or a fixed
scale factor for the end of inflation. From the figure, it is immediate to see how the number of e-folds required to
solve the horizon problem (basically the number of e-folds evolved by our universe since reheating) changes for the
current horizon scale depending on the brane scenario scale  $M_B$. For this figure, we have taken a low inflation
scale $V_*^{1/4} = 10^5$ GeV and the end of inflation to happen at $\ln(a) \approx -40$. Note that the commonly used
value of $N\approx 60$ is obtained for $M_B=1$ GeV.

Before continuing the numerical analysis of the CW brane inflation, we briefly discuss the consistency of introducing a huge change with ${\tilde V}=V/(2\Lambda)\gg 1$ in the Friedman equation.
First, we note that the length scale on the brane is determined by  $L\sim |R_{\mu\nu\alpha\beta}|^{-1/2}\sim |T_{\mu\nu}/M^2_P|^{-1/2}\sim V^{-1/2} M_P$. On the other hand, the length scale in the bulk is related to the AdS length scale, $l\sim \Lambda^{-1/2} M_P$, due to the relations to the input parameters in the RS model, $l=\sqrt{6M^3_5/|\Lambda_{\rm bulk}|}$ and $\Lambda=\sqrt{6M^3_5|\Lambda_{\rm bulk}|}$ with $M^2_P\sim M^3_5\,l$. Thus, taking $V/(2\Lambda)\gg 1$, we are in the regime that $L/l\sim (\Lambda/V)^{1/2}\ll 1$ where the continuum of KK modes in the RS II model with a single brane could be easily excited during inflation and affect the slow-roll inflation on the brane. Therefore, we need to stabilize the radius of extra dimension, for instance, by introducing a second brane in the bulk, so that the KK modes become discrete and decoupled during inflation. We don't go to the details on the radius stabilization in our work and we just assume that a radius stabilization mechanism of Goldberger-Wise type \cite{GW} is at work. Then, the radius should be stabilized at a small value such that the KK masses are larger than the Hubble parameter during inflation. Moreover, in order for the inflaton potential not to destabilize the radius, the mass of the radion, namely, the excitation of the radius, must also satisfy $m^2_r\gtrsim H^2$ where $H^2\sim {\tilde V}_* V_*/M^2_P$. For instance, for $V^{1/4}_*=10^5\,{\rm GeV}$ and $\Lambda^{1/4}=M_B=1\,{\rm GeV}$, we get ${\tilde V}_*\sim 10^{20}$ so the radion mass should be $m_r\gtrsim 100\,{\rm GeV} $.
Consequently, we can accommodate in our setup a light radion relevant for collider physics and still be compatible with inflationary phenomenology contrary to the high-scale case.

\begin{figure}[t]
\centering
\includegraphics[width=7.7cm]{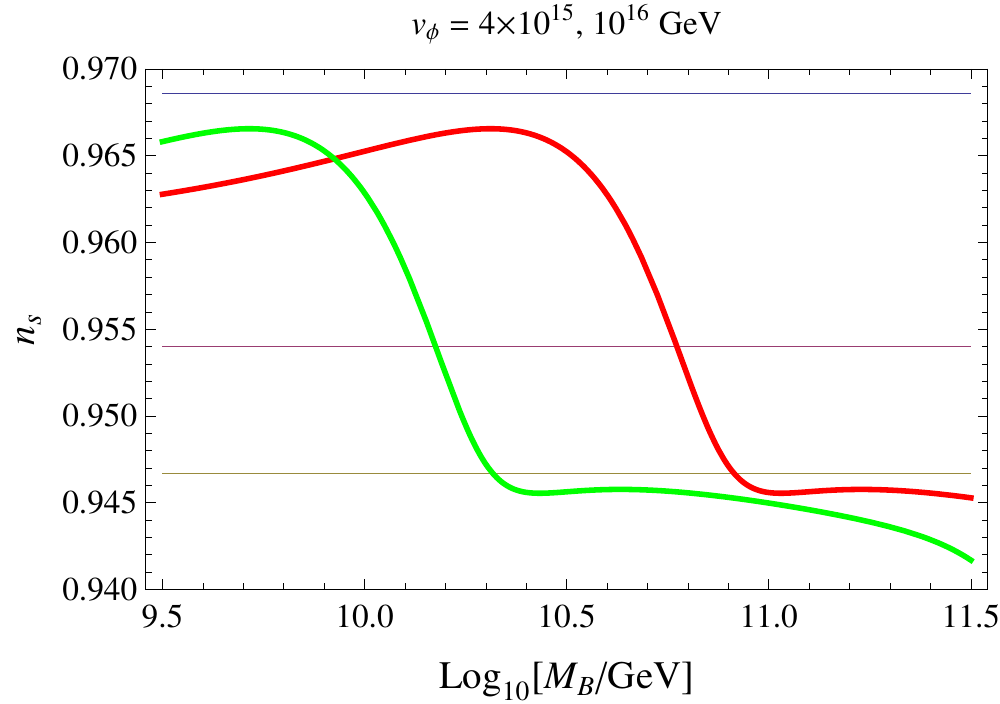} \hspace{1ex}
\includegraphics[width=8.0cm]{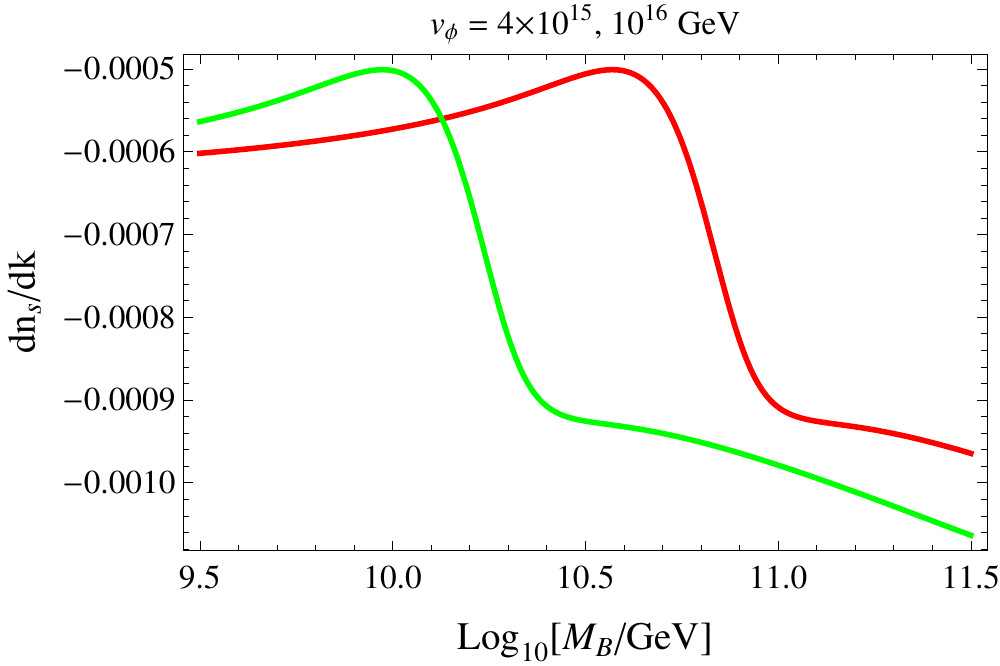} \\[2ex]
\includegraphics[width=8.0cm]{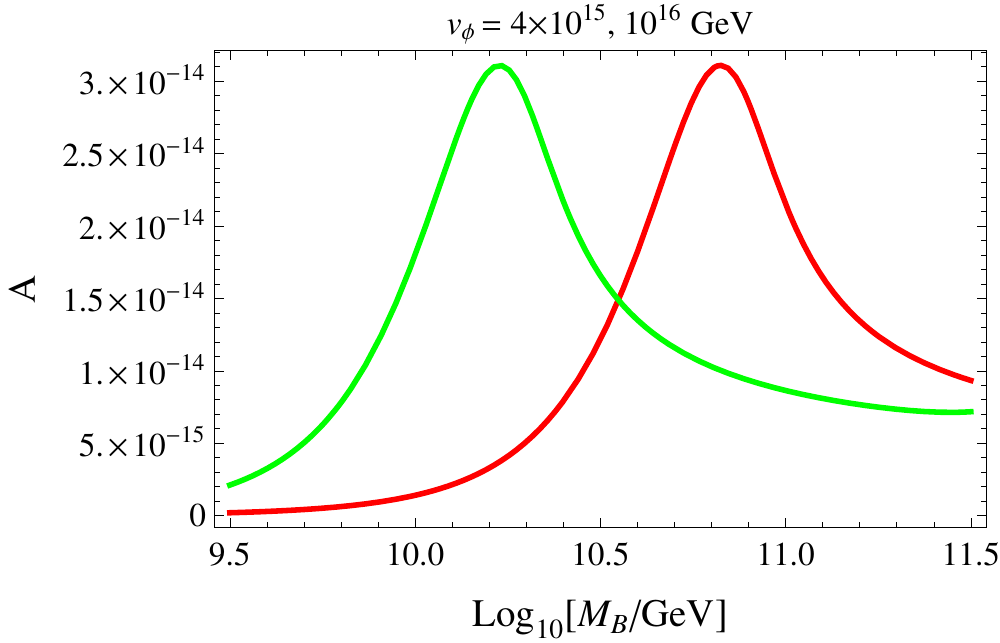} \hspace{1ex}
\includegraphics[width=7.7cm]{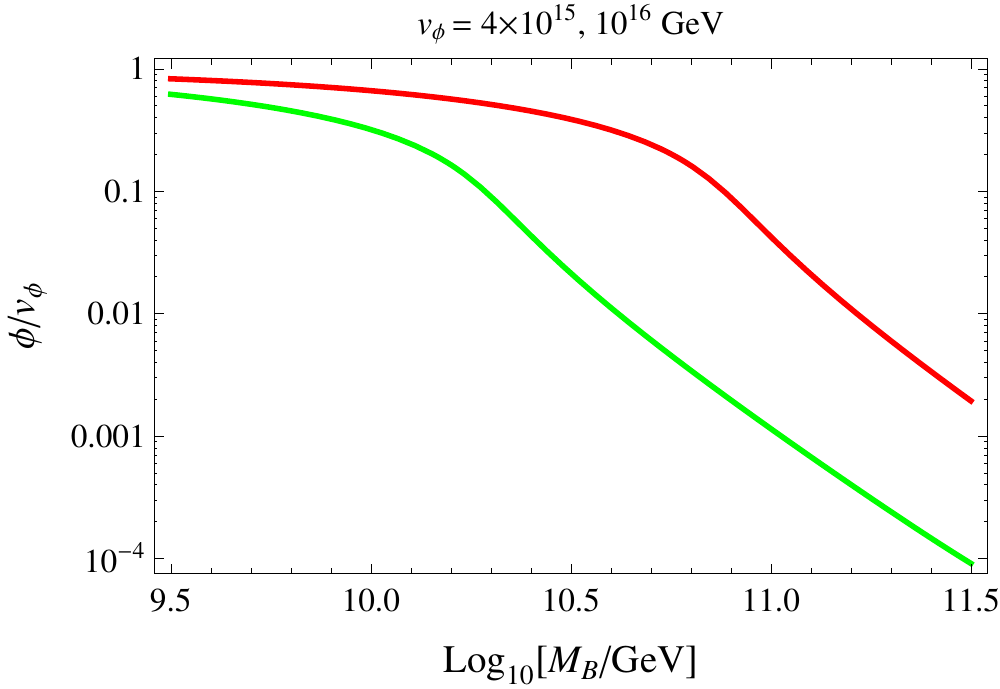}
\caption{The spectral index $n_s$, its running $d n_s/d k$, the quartic coupling $A$ and the horizon-exit field
value of $\phi/v_\phi$ are shown in terms of the brane scale $M_B$  for fixed symmetry breaking scales $v_\phi= 4\times10^{15}$ GeV (left green curve) and $10^{16}$ GeV (right red curve) in each panel. The upper two horizontal
lines in the first panel shows $n_s$ within the $1\sigma$ range of the Planck data and the lower line is
the $2\sigma$ limit.} \label{fig:CWRS16}
\end{figure}

\begin{figure}[t]
\centering
\includegraphics[width=7.7cm]{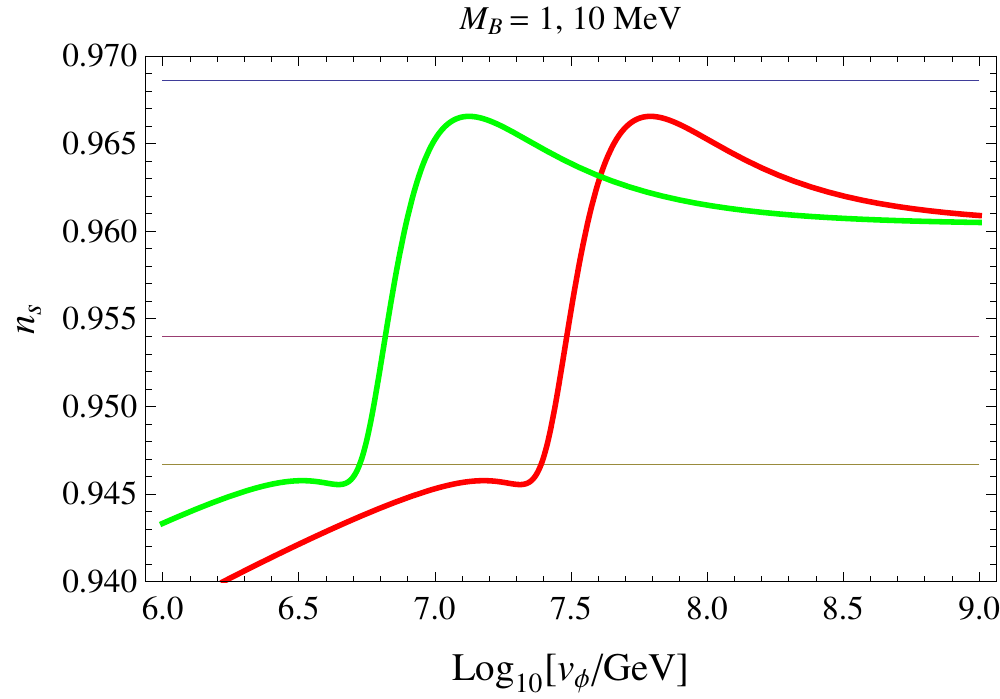} \hspace{1ex}
\includegraphics[width=8.0cm]{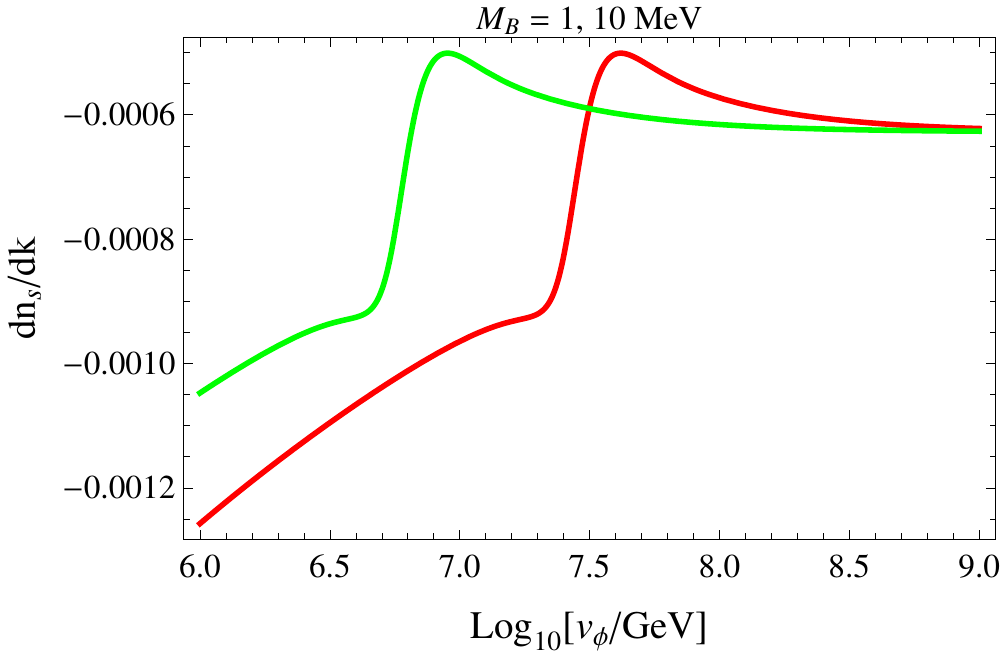} \\[2ex]
\includegraphics[width=8.0cm]{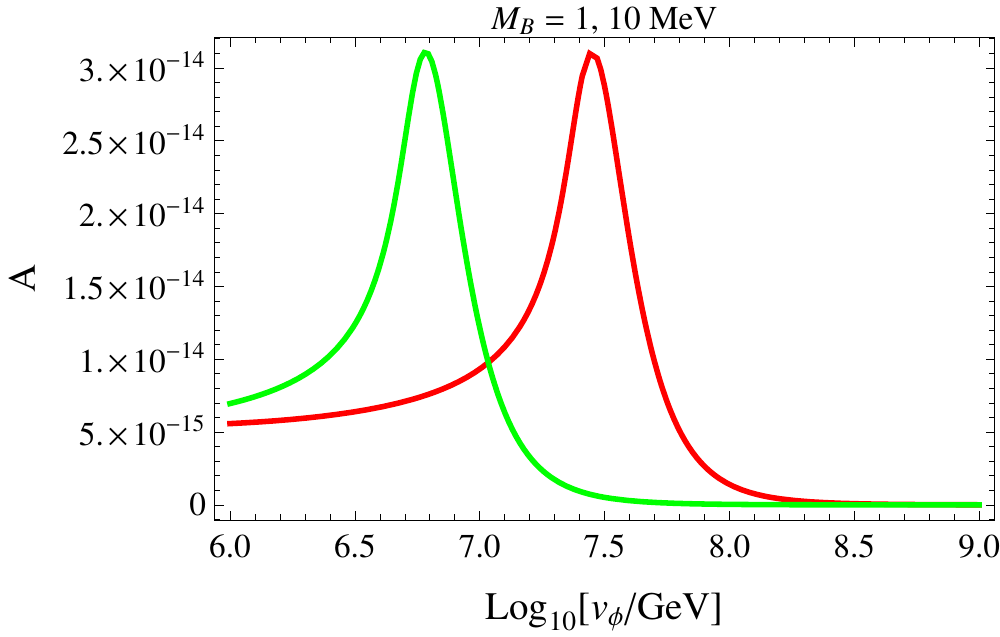} \hspace{1ex}
\includegraphics[width=7.7cm]{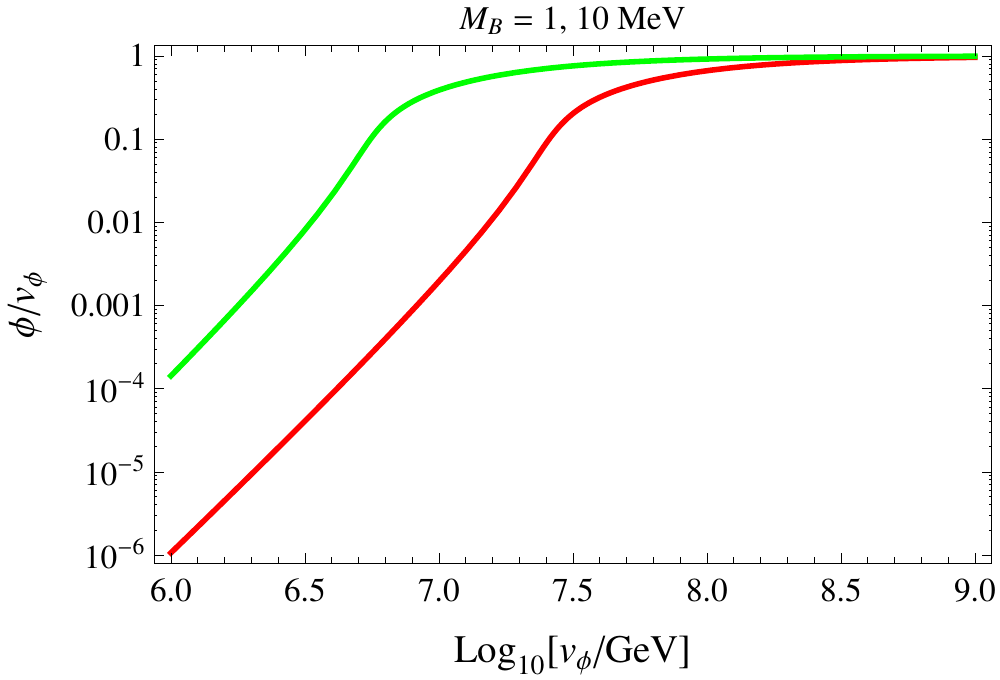}
\caption{The spectral index $n_s$, its running $d n_s/d k$, the quartic coupling $A$ and the horizon-exit field
value of $\phi/v_\phi$ are shown in terms of the symmetry breaking scale  $v_\phi$  for fixed brane scales  $M_B= 1$ MeV (left green curve) and 10 MeV (right red curve) in each panel. The upper two  horizontal lines in the first panel shows $n_s$ within the $1\sigma$ range of the Planck data and the lower line is the $2\sigma$ limit.}
\label{fig:CWRS9}
\end{figure}

Let us now examine how successful inflationary solutions satisfying the conditions,
${\cal P}_{{\cal R},B}(k_*) = 2.2 \times 10^{-9}$ and $N_B= N_{B,*}$, can arise
in the limit of $\tilde V \gg 1$. For given $v_\phi$ and $M_B$, these conditions can be solved uniquely by
appropriate values of $A$ and $\phi/v_\phi$ as shown in Figs.~\ref{fig:CWRS16} (varying $M_B$) and
\ref{fig:CWRS9} (varying $v_\phi$)  for high and low $v_\phi$ ($M_B$), respectively.

When $M_B$ is not so small that the $\phi$-independent prefactor of the modified slow-roll parameters
($e.g.$, $\eta_B \approx \eta/\tilde V$) remains much larger than one:
$(4 M_P/v_\phi)^2 (4 M_B^4 /A v_\phi^4) \gg 1$, one still finds solutions for $\phi/v_\phi \ll 1$ leading to $|\epsilon_B| \ll |\eta_B| \approx \sqrt{2|\xi^2_B|/3}$ as in the standard cosmology. Thus, the simple correlations
in Eqs.~(\ref{Nns}, \ref{ns-run}, \ref{PR}) still hold. This behaviour can be seen from the solution lines
corresponding to the large region of $M_B$ and the small region of $v_\phi$
in each panel of Figs.~\ref{fig:CWRS16} and \ref{fig:CWRS9}, respectively.
As $M_B$ ($v_\phi$) becomes smaller (larger) starting from the right (left) end
in Fig.~\ref{fig:CWRS16} (\ref{fig:CWRS9}), the CW parameters, $\phi/v_\phi$ and $A$, as well as the inflationary observables, $n_s$ and $d n_s /d k$, increase monotonically, and $n_s$ reaches its local maximum at around 0.946 corresponding to $\phi/v_\phi \sim 0.01$.
This is the exactly same pattern as in Figs.~\ref{fig:st1} and \ref{fig:st2} where the four quantities increase as $v_\phi$ gets close to $M_P$, and  $n_s (\phi/v_\phi)$ approaches 0.945 ($0.01$) maximally allowed in the region of $v_\phi < M_P$.
One can also check that the relation (\ref{ns-run}) holds intact
in this small $\phi/v_\phi$ region.
Note that the spectral can be enhanced a lot even for low $v_\phi$ (Fig.~\ref{fig:CWRS9}) by choosing appropriate $M_B$, but is still limited below the Planck's 2$\sigma$ lowest value in the region of $\phi/v_\phi \lesssim 0.01$ as in the standard cosmology.

When $M_B$ ($v_\phi$) becomes even smaller (larger) in Fig.~\ref{fig:CWRS16} (\ref{fig:CWRS9}),
one finds solutions with $\phi/v_\phi \gtrsim 0.01$ eventually approaching the one
for which the standard relations (\ref{Nns}, \ref{ns-run}, \ref{PR}) are invalidated, but new correlations
appear. The spectral index (and also its running) starts to rise rapidly at $\phi/v_\phi\sim 0.1$, but it reaches its maximum value $\sim 0.966$  $(-0.005)$ and then decreases slowly to approach $0.96$  $(-0.006)$. In this large $\phi/v_\phi$ region, the spectral index (and its running) is dominated by
the $\epsilon_B$ contribution.  One can see that there appears a new correlation between these two observables,
which is different from the standard one (\ref{ns-run}).
Note that $A$ becomes much smaller than the typical value $\sim 10^{-14}$ in the asymptotic region preventing the prefactor  $(4 M_P/v_\phi)^2 (4 M_B^4 /A v_\phi^4)$ becoming too small and thus
allowing solutions with $\phi/v_\phi$ close to one.

Let us remark that the four lines in each panel of Figs.~\ref{fig:CWRS16} and \ref{fig:CWRS9}
show the same behavior as functions of  $\phi/v_\phi$, that is, one can always find
an appropriate range of $M_B$ and $v_\phi$ reproducing the same values of $n_s$ and $dn_s/dk$,
as well as the same correlation between them.
Of course, $M_B$ cannot be taken to be smaller than {\cal O}(1) MeV
for which the standard big-bang nucleosynthesis prediction is spoiled.
We find that a spectral index within 1$\sigma$ range of the Planck data can be obtained
for $v_\phi \gtrsim 3 \times 10^7$ GeV with the restriction of $M_B>10$ MeV.
As in the standard cosmology, there is also
a remarkable correlation between the running of the spectral index and the spectral index itself.
Taking the 1$\sigma$ range of the Planck data for the spectral index, $n_s = (0.9540, 0.9686)$,
the CW inflation on the brane predicts
\beq
 {d n_s \over d \ln k} = (-0.00064,-0.0005) .
\eeq
It is worth noticing that this kind of correlation is unique to CW potentials and cannot
be avoided and therefore provides a crucial test on the model.
The ratio of tensor to scalar perturbations is not shown because although it is different from zero and negative, its actual value is so tiny that it is effectively zero from an experimental point of view. Again this feature cannot be circumvented in the CW inflation, and therefore the measurement of gravity waves will completely rule out small-field CW inflation.

The allowed ranges of the symmetry breaking scale,  $v_\phi\sim 10^{8}$ GeV, and the quartic coupling, $A \sim 10^{-14}$, shown in Fig.~\ref{fig:CWRS9}, are of our special interest as such small values of $A$ can
have a dynamical origin in a model where the full scalar potential, including both the inflaton and Higgs sectors, can
be generated purely from radiative corrections.

\section{Dynamical generation of the inflaton potential}

As an specific example of the CW inflation, let us consider the SM extended with a generalized $U(1)_{B-L}$ gauge symmetry.
In its minimal setup where the $U(1)$ charges are given by a linear combination of $B-L$ charge and
hypercharge as $X=Y_{B-L}-x Y$, the model requires three right-handed neutrinos which couple to a $B-L$ Higgs field $\Phi$ and thus
acquire heavy Majorana masses after the $B-L$ symmetry breaking: \beq {\cal L}_{B-L} = - y_N \Phi \overline{(\nu_R)^c}
\nu_R \,. \label{yukawas} \eeq The scalar potential at tree-level is written as \beq V_{tree}(H,\Phi) = m_H^2 |H|^2 +
m_S^2 |S|^2 +\lambda_H |H|^4 + \lambda_\Phi |\Phi|^4 + \lambda_{H\Phi} |H|^2 |\Phi|^2 \eeq where $H$ denotes the SM
Higgs field. As noted in Ref.~\cite{chun13}, all of the above Higgs potential parameters, consistent with the recent
SM Higgs data \cite{lhc}, can be generated radiatively assuming a vanishing initial condition at a certain high scale $\Lambda_{ UV}$. All the scales are generated by dimensional transmutation of the CW mechanism applied to the $B-L$ gauge
symmetry breaking. In this setup, there are two free parameters, the extra gauge coupling $g_X$ and the
right-handed neutrino Yukawa coupling $y_N$, from which the $B-L$ breaking scale $v_\phi$ as well as the $B-L$ Higgs
quartic coupling $\lambda_\Phi$ are generated dynamically. Extending the analysis of Ref.~\cite{chun13} to higher
$v_\phi$ scale, we will examine whether there exists an appropriate CW minimization point which is consistent with the
observed cosmological quantities derived in the previous section.

\medskip

\begin{figure}[t]
\centering
\includegraphics[width=10cm]{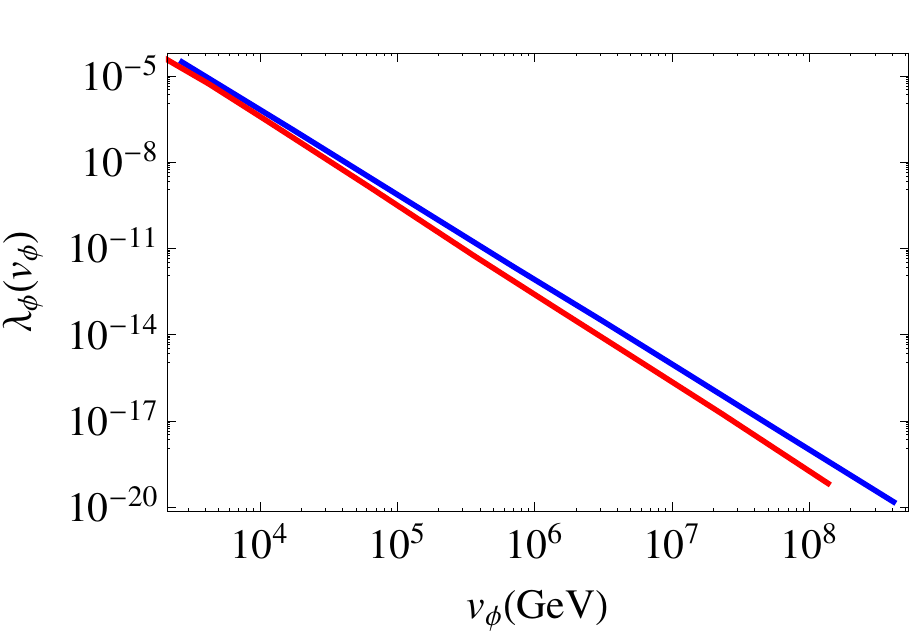}
\caption{  The values of $v_\phi$ vs.\ $\lambda_\Phi$ satisfying the CW minimization condition \cite{chun13}. The upper (blue), and lower (red) lines correspond to the UV scale $M_X=2\times 10^{11}$ GeV, and $10^{18}$ GeV, respectively. } \label{fig:BL}
\end{figure}

Let us now consider the one-loop Coleman-Weinberg potential \cite{CW} for the $B-L$ sector. Taking
$\Phi=\phi/\sqrt{2}$ in the unitary gauge and the normalization condition of $V''(0)=0$ and $V''(Q)=6 \lambda_\Phi$,
the one-loop corrected $B-L$ potential is given by \cite{CW}
\beq V_{X}(\phi)=\frac{1}{4}\lambda_\Phi \phi^4
+\frac{1}{64\pi^2} \phi^4 \Big(10\lambda^2_\Phi+48 g^4_X - 8y^4_{N}\Big)\bigg(\ln
\frac{\phi^2}{Q^2}-\frac{25}{6}\bigg) +V_0 \label{CWpot} \eeq where we took one Yukawa coupling for the right-handed
neutrinos $y_N$, and added a constant term $V_0$ normalizing the potential: $V_X=0$ at the global minimum. Taking
the renormalization scale at $Q=\langle\phi\rangle\equiv v_\phi$ to avoid the large-log uncertainty in the one-loop
approximation \cite{CW}, one can evaluate the minimization condition of the potential  (\ref{CWpot}) and obtain \beq
\lambda_\Phi(v_\phi)=\frac{11}{48\pi^2}\Big(10\lambda^2_\Phi+48 g^4_X -8 y^4_{N}\Big)(v_\phi).
\label{Vmin} \eeq This relation fixes the $B-L$ breaking scale $v_\phi$ in terms of input values of $\lambda_\Phi$,
$g_X$ and $y_N$ which evolve from the high scale $\Lambda$ to $v_\phi$ by renormalization group. Putting back
(\ref{Vmin}) into (\ref{CWpot}), one has \beq V_X(\phi) = {3 \lambda_\Phi \over 22} \phi^4 \left( \ln \left( \phi
\over v_\phi \right) -{1\over4} \right) + {3 \lambda_\Phi\over 88} v_\phi^4 , \eeq which is nothing but the general CW
potential given in Eq.~(\ref{Apot}) with $\lambda_\Phi \equiv  22 A/3$.

In Figure~\ref{fig:BL} we present the values of $v_\phi$ and $\lambda_\Phi$ satisfying the CW minimization condition as well as the correct electroweak symmetry breaking for appropriate values of $g_X$. As noted in Ref.~\cite{chun13}, one can see that larger $v_\phi$ requires smaller $g_X$.  From Figure~\ref{fig:CWRS9}, one can see that the spectral index $n_s$ falls into the 1$\sigma$ range of the Planck data for $v_\phi \gtrsim 5\times10^6\,{\rm GeV}$ which requires $\lambda_\Phi \approx 22 A/3 \lesssim 2\times10^{-13}$.
Furthermore, the required values of $\lambda_\Phi$ drop rapidly for higher $v_\Phi$.
Thus one finds viable parameter points around $v_\phi\sim 10^8$ GeV and $M_B\sim 1$ MeV 
in which the inflaton and Higgs potentials
are generated simultaneously purely from radiative corrections. 
However, we should remark that the solutions to the
minimization condition, $\lambda_\Phi \sim 11/\pi^2( g^4_X - y_N^4/6)$, are found for fine-tuned  choices of  $ g_X^4\simeq  y_N^4/6$ resulting in highly suppressed values of $\lambda_\Phi$.
It is expected that much less fine-tuned solutions would be found in some other $U(1)^\prime$ models which have different beta-function coefficients.

\section{Conclusions}
We have analyzed the plausibility of a (small field) CW potential naturally arising from quantum corrections to
address inflation in and beyond the standard cosmological scenario. We have shown that although it is not possible in
the standard scenario to solve the horizon problem within $1\sigma$ consistency with the Plank measurement of the
spectral index, brane-world scenarios ease the requirements on the numbers of e-folding needed to solve the horizon
problem, while preserving the correlations among the other observables, allowing   CW potentials to fulfill all the
inflation requirements and remain an attractive and natural explanation for inflation. Besides, CW potentials have
an inherent prediction on the relation between the running of the spectral index and the spectral  index itself,
which can be tested in the near future. Tensor modes are essentially absent in  CW inflation models and thus,
if found by next generation experiments, can rule out CW motivated inflationary potentials altogether.

The inflation observables can easily accommodate a rather small symmetry breaking scale  $\sim 10^8$ GeV but do require a tiny quartic coupling $\sim 10^{-18}$. While the CW mechanism generates small scales  through dimensional
transmutation, such a tiny coupling may be indicative of a radiative origin. We have illustrated how dynamical
generation of the inflaton as well as the SM Higgs potential can work consistently with the Planck results as well as
the Higgs mass measurement at the LHC in the context of the SM extended with the generalized $B-L$ gauge symmetry to explain the
neutrino masses and mixing.

\section*{Acknowledgments}
G.B wish to warmly thank KIAS Physics Group for its hospitality. She acknowledges support from the MEC and FEDER (EC)
Grants FPA2011-23596 and the Generalitat Valenciana under grant PROMETEOII/2013/017. HML was supported in part by
Basic Science Research Program through the National Research Foundation of Korea (NRF) funded by the Ministry of Education, Science and Technology (2013R1A1A2007919). EJC was supported by SRC program of NRF Grant No.\ 2009-0083526
funded by the Korea government (MSIP) through Korea Neutrino Research Center.

\end{document}